\documentclass{article}
\usepackage{spconf}

\usepackage{amsmath,amsfonts, amsthm, bm, amssymb}

\def\Figref#1{Figure~\ref{#1}}

\def\secref#1{section~\ref{#1}}

\def\eqref#1{equation~(\ref{#1})}

\def\plainref#1{(\ref{#1})}

\def\1{\bm{1}}

\def\vtheta{{\bm{\theta}}}

\def\vphi{{\bm{\phi}}}

\def\ve{{\bm{e}}}
\def\vf{{\bm{f}}}

\def\vx{{\bm{x}}}

\def\mD{{\bm{D}}}

\def\mK{{\bm{K}}}
\def\mL{{\bm{L}}}

\def\mT{{\bm{T}}}

\def\mW{{\bm{W}}}

\def\mPhi{{\bm{\Phi}}}

\DeclareMathAlphabet{\mathsfit}{\encodingdefault}{\sfdefault}{m}{sl}
\SetMathAlphabet{\mathsfit}{bold}{\encodingdefault}{\sfdefault}{bx}{n}

\def\gG{{\mathcal{G}}}

\def\gO{{\mathcal{O}}}

\def\gW{{\mathcal{W}}}

\def\sE{{\mathbb{\xi}}}

\def\sP{{\mathbb{P}}}

\def\sS{{\mathbb{S}}}

\def\sV{{\mathbb{V}}}

\newcommand{\R}{\mathbb{R}}

\usepackage[ruled, lined, linesnumbered]{algorithm2e}
\usepackage{url}
\usepackage{xcolor}
\usepackage[colorlinks = true,
            linkcolor = blue,
            urlcolor  = blue,
            citecolor = blue,
            anchorcolor = blue]{hyperref}

\usepackage{enumitem}
\usepackage{rotating}
\usepackage{tabulary}
\usepackage{multirow}

\usepackage{graphicx, epstopdf}
\usepackage{wrapfig}
\usepackage{caption}
\usepackage{subcaption}
\graphicspath{{./figs/}}

\newtheorem{lemma}{Lemma}
\newtheorem{proposition}{Proposition}

\usepackage{booktabs} 
\title{Efficient graph construction for image representation}
\name{Sarath Shekkizhar, Antonio Ortega\thanks{Our work was supported by DARPA's LwLL program (FA8750-19-2-1005). Code for the proposed method is   available at \href{https://github.com/STAC-USC/NNK_Image_graph}{github.com/STAC-USC/NNK\_Image\_graph}}}
\address{University of Southern California \\
Los Angeles, CA, USA \\
shekkizh@usc.edu, ortega@sipi.usc.edu}
\begin{document}
\ninept
\maketitle
\begin{abstract}
Graphs are useful  to interpret widely used image processing methods, e.g., bilateral filtering,  or  to develop new ones, e.g., kernel based techniques. 
However, simple graph constructions are often used, where edge weight and connectivity depend on a few parameters. In particular, the sparsity of the graph is determined by the choice of a window size. 
As an alternative, we extend and adapt to images recently introduced non negative kernel regression (NNK) graph construction. In NNK graphs sparsity adapts to intrinsic data properties. Moreover, while previous work considered NNK graphs in generic settings, here we develop novel algorithms that take advantage of image properties, so that the NNK approach can scale to large images. 
Our experiments show that sparse NNK graphs achieve improved energy compaction and denoising performance when compared to using graphs directly derived from the bilateral filter. 
\end{abstract}
\section{Introduction}
\label{sec:intro}
A recent trend in image processing has been to move from simple non adaptive filters to   image dependent filters such as the bilateral filter \cite{TomasiBF1998}, non local means \cite{BuadesNLM2005}, block matching \cite{BM3D_2007}, kernel regression methods \cite{KernelRegressionImageProcessing}, expected patch likelihood maximization (EPLL) \cite{ZoranEPLL_2011} or window nuclear norm minimization (WNNM) \cite{GuWNNM_2014}. 
One shortcoming of these adaptive filters is that they cannot be efficiently described using traditional image domain Fourier techniques, since these models are highly non linear. To solve this issue of interpretability, graph based perspectives have been introduced to analyze  data dependent image processing models \cite{milanfar2013tour, PeyreSpectralUnderstanding}. 
In the graph formulation, pixels corresponds to the nodes of a graph and are connected with edges having edge weights capturing pixel similarity. 
\cite{milanfar2013tour, AkshayBF2013} shows the correspondence between window based filters and graph based filtering where the graph is formed by connecting each pixel (node) to only those within a window centered at the pixel.
This method of graph construction for $w \times w$ filtering resembles a $K$-Nearest Neighbor (KNN) graph with $K = w^2$ where neighbors are selected based on their spatial location relative to the pixel at the window center.
Thus, the choice of $w$, similar to  $K$ in KNN graphs,  can be considered a hyperparameter offering coarse control of sparsity and complexity of the graph representation.

In this work, we focus on graph construction for image representation, a specific application which is often overlooked in data driven graph learning methods \cite{Qiao2018}.
Our proposed method extends to images recently introduced  non negative kernel regression (NNK) method \cite{shekkizhar2019graph, shekkizhar2020graph} for graph construction.   
Unlike earlier methods, our framework leads to a principled way to construct sparse graphs that can be combined with spectral graph wavelets and other graph signal processing tools \cite{ortega2018, GeneGraphImage2018}. Relative to previous work  \cite{shekkizhar2019graph, shekkizhar2020graph}, the key novelty in this paper is to exploit pixel position regularity, and specific characteristics of kernels used for image filtering, to learn image graphs in a fast and efficient manner, which allows us to scale the proposed methods to graphs with millions nodes as required by image processing applications.  Experimentally, our image specific simplifications lead to average speed ups in graph construction of at least a factor $10$ for $w=11$, relative to the original NNK algorithm. 

We focus our graph construction and filtering by taking the bilateral filter graph as starting point, but the same can be adapted to other  image processing models that have a graph interpretation (e.g., those described in \cite{milanfar2013tour}).
Of particular relevance to our work is \cite{AkshaySparseBF2017}, where the authors construct sparse graph alternatives by approximating the inverse of the bilateral filter matrix. The authors motivate the idea by drawing parallels to the graph construction methods that estimate sparse inverse covariance or precision matrix of a Gaussian Markov Random Field model \cite{HilmiGraphLaplacian}. The authors note that this method is expensive and resort to a heuristic algorithm which still can only be used for small images \cite{AkshaySparseBF2017}. In contrast, in this paper we are able to apply our method to images with typical sizes. 

We combine our method  with Spectral Graph Wavelets \cite{HammondSGW2011} to illustrate its benefits for image representation, showing that our graphs have $90\%$ fewer edges than bilateral filter graphs constructed with $11\times11$ window while offering better low frequency representation and improved performance in the context of a simple denoising task. The runtime of graph wavelets and other graph filter operations for images are notably reduced due to the sparse nature of our graphs (e.g., $15\times$ faster than the same algorithm using BF graph). 
\section{Preliminaries}
\label{sec:preliminaries}
\subsection{Bilateral Filter}
The bilateral filter (BF) can be interpreted as a \textit{graph filter} on a dense, image-dependent graph, 
with edge weights between nodes (pixels) $i$ and $j$ given by the kernel
\begin{align}
    \mK_{i,j} &= \exp\left(-\frac{\|\vx_i - \vx_j\|^2}{2\sigma_d^2 }\right)\exp\left(-\frac{\|\vf_i - \vf_j\|^2}{2\sigma_f^2 }\right) \label{eq:bf_kernel}
\end{align}
where $\vx_i$ and $\vf_i$ denote the position and the intensity of pixel $i$, respectively. The bilateral filter operation on graph signal $\vf$ can be interpreted as $\mD^{-1}\mK\vf$, where $\mD$ is the degree matrix of the graph and its inverse is used as a normalization. With this interpretation it is also possible to develop alternative graphs, via symmetrization of $\mD^{-1}\mK$  \cite{MilanfarSymmetrizingFilter2013} or sparsification of the original BF graph \cite{AkshaySparseBF2017}. 
Note that the pixel positions $\vx$ and the distances $\|\vx_i - \vx_j\|^2$ in the BF kernel \plainref{eq:bf_kernel} are image independent and are known in advance. 

\subsection{Non Negative Kernel regression graphs}
Note that $\mK_{i,j}$ can be viewed as the inner product of two kernel functions $\vphi_i$ and  $\vphi_j$ \cite{Hofmann2008}.  
Then, a non negative kernel regression (NNK) graph can be computed by formulating graph construction  as a signal representation problem,  where each node is to be approximated by a weighted sum of functions from a dictionary formed by its neighboring nodes  \cite{shekkizhar2019graph,shekkizhar2020graph}. 
At each node, we need to solve:
\begin{align}
& \min_{\vtheta \colon \vtheta \geq 0} \;
||\vphi_i - \mPhi_S\vtheta||^2_{_{l_2}}, \label{eq:nnk_lle_objective}
\end{align}
where $\vphi_i$ (corresponding to node $i$) is to be represented by a linear combination of atoms~(with weights given by $\vtheta$) from a dictionary obtained from a set $S$ of neighbors~($\mPhi_S$).
As shown in \cite{shekkizhar2019graph,shekkizhar2020graph}, it is possible to associate a geometric interpretation to conditions that determine whether two nodes are connected (kernel ratio interval or KRI conditions). Thus, in contrast to KNN graphs, even when window size $w$ increases the number of connected nodes does not necessarily grow, so that NNK graphs tend to be better at reflecting the actual data topology. 

\begin{figure}[htbp]
\centering
\includegraphics[width=0.3\textwidth]{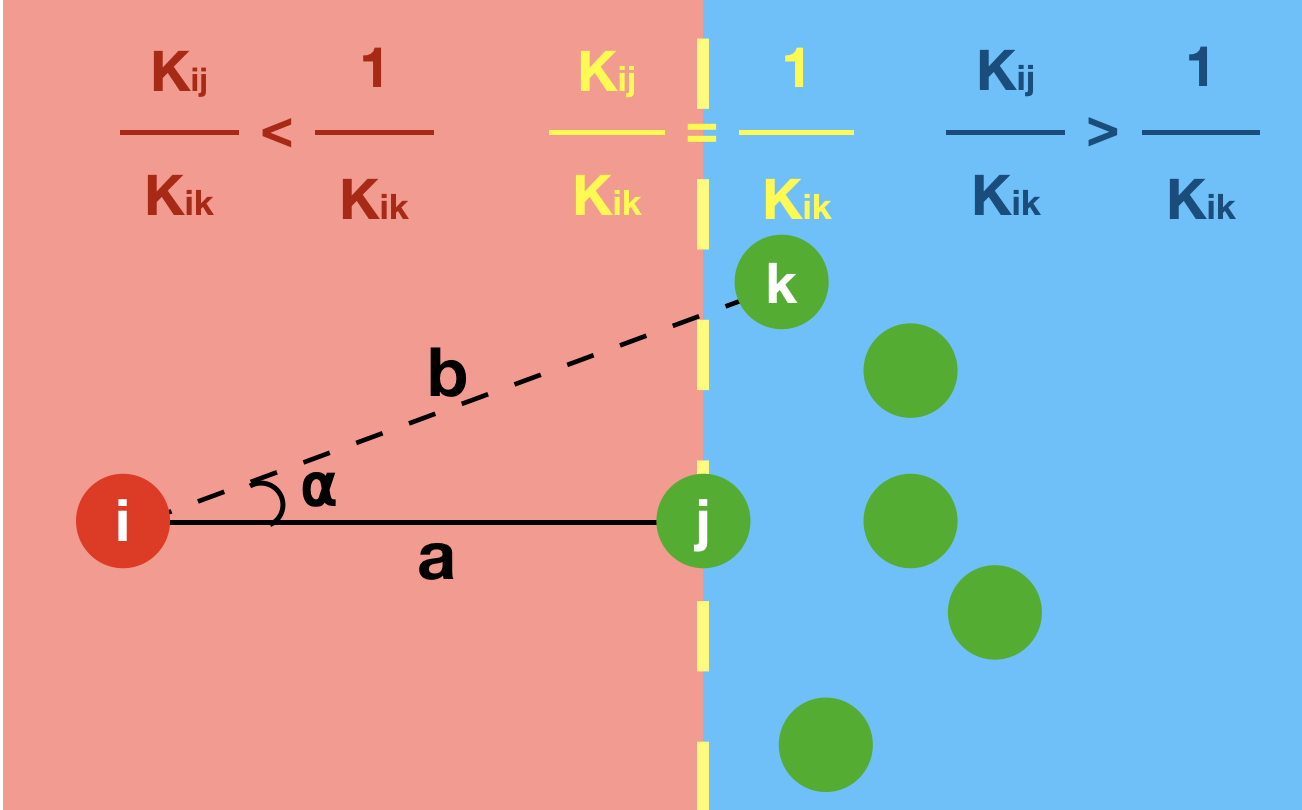}
\caption{Plane (yellow dashed line) associated to a connected node $j$ in NNK. Nodes in the blue region to the right of the plane will be not be connected to node $i$. This provides an intuitive explanation of KRI:  given that there is  edge between node $i$ and $j$, a farther away node $k$ in the same direction can be connected to $i$ only if it is \emph{more similar} to $i$ than $j$.}
\label{fig:plane}
\end{figure}

A key contribution of this paper is to adapt NNK to image data, and in particular taking into account the special characteristics of image to apply the KRI conditions more effectively. Unlike in the general case, where all data dimensions are irregular and unknown, in image graphs pixel locations are regularly spaced and are known before hand. This observation allows us to reduce the KRI condition to simple intensity thresholding rules for removing neighbors in images (see \secref{sec:nnk_image}). 
Formally, the KRI theorem states that for any positive definite kernel with range in $[0, 1]$ (e.g. bilateral kernel \plainref{eq:bf_kernel}), the necessary and sufficient condition for two nodes $j$ and $k$ to be {\em both} connected to node $i$ in a NNK graph is
\begin{align}
\mK_{j,k} < \frac{\mK_{i,j}}{\mK_{i,k}} < \frac{1}{\mK_{j,k}}. 
\label{eq:kernel_ratio_interval}
\end{align}
The geometric interpretation of (\ref{eq:kernel_ratio_interval}) is illustrated by \Figref{fig:plane}.  

\subsection{Spectral Graph Wavelet Transform}
Given the adjacency matrix $\mK$ of a graph $\gG= (\sV, \sE)$, the associated combinatorial graph Laplacian $\mL$ is defined as $\mL = \mD - \mK$, where $\mD$ is the diagonal degree matrix given by $\mD_{i,i} = \sum_j \mK_{i,j}$. A graph signal is a function $f \colon \sV \to \R$ defined on the vertices of the graph. In the case of images, this corresponds to the intensity or values defined at each pixel. 
The Graph Fourier Transform (GFT) \cite{ortega2018}  is defined as the expansion of the graph signal in terms of the eigenvectors of chosen graph operator, e.g., the graph Laplacian.

Spectral Graph Wavelets (SGW) \cite{HammondSGW2011} are based on defining a scaling operator in the GFT domain, based on the eigenvectors of the graph, analogous to the Fourier transform but for signals defined on a weighted graph. 
The wavelet coefficients for a given signal at scale $s$ at a vertex $i$ are calculated as a function of a graph operator $\mT_g = g(\mL)$ and the eigenpair ($\lambda_l, \ve_l$) defining the GFT, namely
\begin{align}
\gW_f(s, i) = \mT^s_gf(i) = \sum^{|\sV|}_{l=1}g(s\lambda_l)\hat{f}(\lambda_l)\ve_l(i) \label{eq:wavelet_coefficents}
\end{align}

These coefficients can be computed with a fast algorithm based on Chebychev polynomials for approximating the scaled operator function. We refer the reader to \cite{HammondSGW2011} for further details on approximations for practical realization of SGW.

\section{NNK IMAGE FILTERING}
\label{sec:nnk_image}
Given the neighbor set at each node, NNK graph can be obtained with $\gO(K^3)$ complexity at each node, where $K$ is the number of neighbors. 
In this section, we present image specific simplifications to compute NNK graphs efficiently.

\subsection{Kernel Ratio Interval for images}
The KRI condition of \plainref{eq:kernel_ratio_interval} allows us to identify neighboring nodes (pixels in $w \times w$ window for images) that will have zero edge weights, given knowledge of a connected node (\Figref{fig:plane}).
From an image point of view, this corresponds to removing graph edges 
to pixels which are farther away when there exists pixels with similar intensity that are closer. 
\begin{proposition}
\label{prop:kri_images}
The necessary and sufficient condition for a pixel $k$ to not have an edge to pixel $i$ given that pixel $j$ is connected to pixel $i$ i.e $\vtheta_{i,k} = 0 | (\vtheta_{i,j} > 0)$ is given by
\begin{equation}
    (\vf_j - \vf_k)^\top(\vf_j - \vf_i) < \left(\frac{\sigma_f}{\sigma_d}\right)^2(\vx_k - \vx_j)^\top(\vx_j - \vx_i)
    \label{eq:proposition-1}
\end{equation}

\begin{proof}
Denote $d_{i,j} = \|\vx_i - \vx_j\|^2$ and similarly $f_{i,j} = \|\vf_i - \vf_j\|^2$. Thus the bilateral filter weights can be rewritten as:
\begin{align}
    \mK_{i,j} 
    &= \exp\left( - \frac{d^2_{i,j}}{2\sigma_d^2 } - \frac{f^2_{i,j}}{2\sigma_f^2 }\right) \label{eq:simplified_bf_kernel}
\end{align}
Now, the contraposition of KRI theorem \plainref{eq:kernel_ratio_interval} gives a necessary and sufficient condition for an edge ($\vtheta_{i,k}$) to be disconnected as
\begin{align}
    \frac{\mK_{i,j}}{\mK_{i,k}} & \geq \frac{1}{\mK_{j,k}}. \label{eq:kri_contrapositive}
\end{align}
Substituting for the bilateral weight kernel
\begin{align}
   \frac{\mK_{i,j}}{\mK_{i,k}} 
   & = \exp\left( -\frac{(d^2_{i,j} - d^2_{i,k})}{2\sigma^2_d} -\frac{(f^2_{i,j} - f^2_{i,k})}{2\sigma^2_f} )\right) \label{eq:kri_image_left}
\end{align}
Thus, condition \plainref{eq:kri_contrapositive} is simplified as
\begin{align*}
    \exp\left( -\frac{(d^2_{i,j} - d^2_{i,k})}{2\sigma^2_d} -\frac{(f^2_{i,j} - f^2_{i,k})}{2\sigma^2_f})\right) \geq  \exp\left(\frac{d^2_{j,k}}{2\sigma_d^2 } + \frac{f^2_{j,k}}{2\sigma_f^2 }\right) 
\end{align*}
Taking logarithm on both sides and rearranging terms corresponding to intensity and location we obtain after some manipulations
\begin{align}
     f^2_{i,j} + f^2_{j,k} - f^2_{i,k}  & \leq - \left(\frac{2\sigma^2_f}{2\sigma^2_d}\right) d^2_{i,j} + d^2_{j,k} - d^2_{i,k} \label{eq:kri_simplified_1}
\end{align}
Using the simplification from Lemma \ref{lemma:norm_simplification}, to replace $d^2_{i,j} + d^2_{j,k} - d^2_{i,k}$ and $f^2_{i,j} + f^2_{j,k} - f^2_{i,k}$  leads to (\ref{eq:proposition-1}) and concludes the proof. 
\end{proof}
\end{proposition}
\begin{lemma}
\label{lemma:norm_simplification}
\begin{align*}
   d^2_{i,j} + d^2_{j,k} - d^2_{i,k}  &= 2\;(\vx_j - \vx_k)^\top(\vx_j - \vx_i) \\
   f^2_{i,j} + f^2_{j,k} - f^2_{i,k}  &= 2\;(\vf_j - \vf_k)^\top(\vf_j - \vf_i)
\end{align*}
\begin{proof}
Omitted for space.
\end{proof}
\end{lemma}

\subsection{Sparse graph representation of images}
\begin{figure}[htbp]
    \centering
    \begin{subfigure}{0.21\textwidth}
    \centering
    \includegraphics[width=\textwidth]{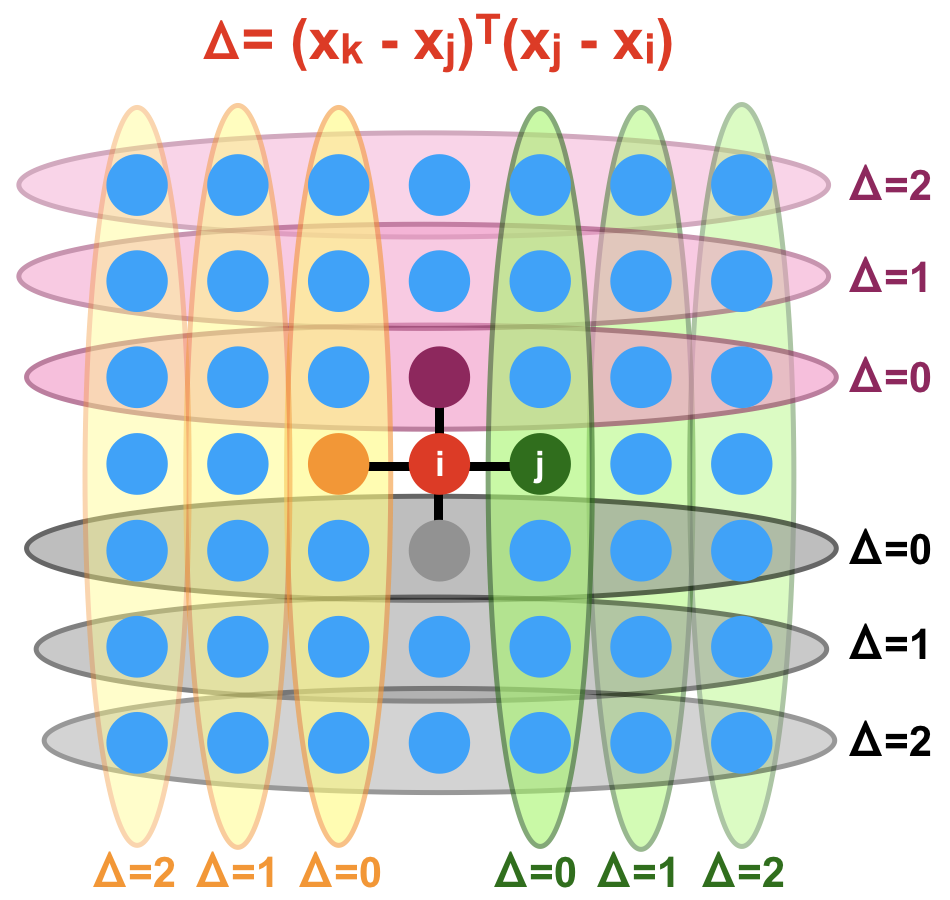}
    \caption{}
    \label{fig:kri_for_images}
    \end{subfigure}
    \begin{subfigure}{0.23\textwidth}
    \centering
    \includegraphics[trim={0cm 6cm 2cm 6cm},clip,width=\textwidth]{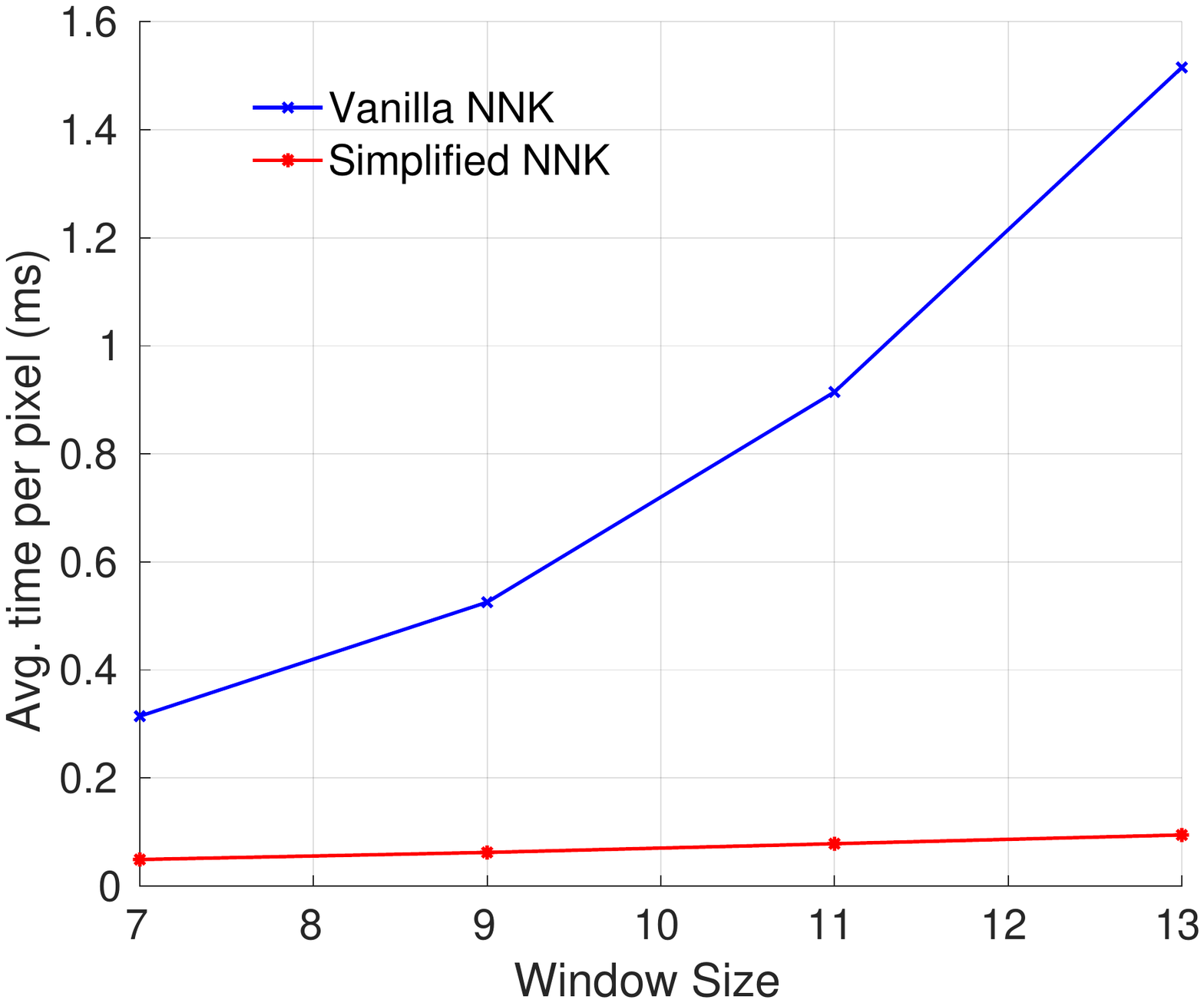}
    \caption{}
    \label{fig:runtime_difference}
    \end{subfigure}
    \caption{\textbf{(a)}. (Best viewed in color) Simple scenario of 4 connected neighbor and remaining pixels in a $7\times7$ window with their associated \emph{threshold factor} ($\Delta$). For e.g, given pixel $j$ is connected to $i$, the proposed graph construction eliminates all pixel intensities in green region (right of $i$) which satisfy the condition in  Proposition \ref{prop:kri_images}. 
    The algorithm continues pruning by moving radially outwards connecting pixels that are not pruned and removing ones that are to be pruned based on proposition until no pixel is left unprocessed.\\
    \textbf{(b)}. Average processing time per pixel for our proposed simplified NNK and the original NNK construction. We observe a similar trend on all our test images with the difference widening further for increasing window sizes.}
\end{figure}

The right term in (\ref{eq:proposition-1}) depends only on pixel locations and kernel parameters and can be determined for a given window size $w$ and saved before hand. As a further simplification we consider only threshold factors ($\Delta = (\vx_k - \vx_j)^\top(\vx_j - \vx_i)$) that are positive, corresponding to regions along the same direction as the connected pixel $j$ (see \Figref{fig:kri_for_images}). This stands to intuition as the KRI plane (\Figref{fig:plane}) would hardly influence the selection of pixels on the other side of the window. The order of the pixels in the window can be precomputed by closest to farthest for a given window.
Further, given the set of neighbors for each pixel after pruning, we approximate the weights with the original bilateral kernel weights, instead of computing them to satisfy the condition of (\ref{eq:nnk_lle_objective}). This can be justified by the fact that both NNK and original BF kernel weights maintain the same relative order of importance and would serve as good approximations.
Let us consider the simple case where only two nodes remain after pruning, then NNK weights are given as
\begin{align*}
    \begin{bmatrix}
        \vtheta_{i,j} \\ \vtheta_{i,k}
    \end{bmatrix} = \frac{1}{1-\mK^2_{j,k}}
    \begin{bmatrix}
        \mK_{i,j} - \mK_{j,k}\mK_{i,k} \\ \mK_{i,k} - \mK_{j,k}\mK_{i,j}
    \end{bmatrix}
\end{align*}
Thus,
\begin{align}
    \vtheta_{i,j} - \vtheta_{i,k} = (\mK_{i,j} - \mK_{i,k})\left[ \frac{1+\mK_{j,k}}{1-\mK^2_{j,k}}\right]
\end{align}
The factor on the right with $\mK_{j,k}$ term is strictly positive and thus the relative impact of the edges is preserved when NNK graph is approximated with the original kernel weights.
\begin{algorithm}
\DontPrintSemicolon
\SetKwInOut{Input}{Input}
\SetKwFunction{Precompute}{Precompute}
\SetKwFunction{FMain}{NNK\_Image\_Graph}
\SetKwProg{Fn}{Function}{:}{}
\Fn{\Precompute{$w$}}{
$\vx = \text{pixel positions in } w\times w,\;\; \sS = \{ \text{pixels in } w\times w \}$\\
window center $= i,\;\; \sS = \sS - \{i\}$\\
$\sS = $ sort $\sS$ by $ \|\vx_j - \vx_i\|^2 \;\; \forall j \in \sS$ \\
\For{$j,k$ in $\sS$}{
        $\Delta_{j,k} = (\vx_k - \vx_j)^\top(\vx_j - \vx_i)$
}
}
\textbf{return} Ordered window pixels $\sS$, Threshold factor $\Delta$ \\[0.5em]
\Input{Image $\vf \in  \R^{m\times n}$, window size $w, \sigma_f, \sigma_r$}
\Fn{\FMain{}}{
$\mu = \left(\frac{\sigma_f}{\sigma_d}\right)^2$,  \hspace{0.5em}
$[\sS^*,\; \Delta ]= \;$\Precompute($w$)\\
    \For{ each pixel $i$}{
        $\sS = \sS^*,\;\; \sP = \{\}$ \\
        \For{$j$ in $\sS$}{
        \tcc{pick neighbors in spatial distance sorted order}
        \uIf{$j$ in $\sP$}{continue \tcp*[r]{skip if pruned}}
        $\sP = \sP + \{j\}$ \\
        \For{pixel $k$ in $\sS$ with $\Delta_{j,k} \geq 0$}{
        \tcc{consider pixels in same direction as $j$}
            \uIf{$(\vf_j - \vf_k)^\top(\vf_j - \vf_i) \leq \mu\Delta_{j,k} \;$}{$\sP = \sP + \{k\}, \;\; \sS = \sS - \{k\}$}
            }
        }
        $\mW_{i, \sS} = \mK_{i, \sS}\;, \;\;\mW_{i, \sS^c} = 0 $
    }
}
\textbf{return} Graph Adjacency $\mW$
\caption{Proposed NNK for Images}
\label{algorithm:nnk_algorithm}
\end{algorithm}

\section{IMAGE REPRESENTATION AND DENOISING WITH NNK IMAGE GRAPHS}
\label{sec:experiments}
We validate experimentally the effectiveness of our proposed method over the naive BF graph version in terms of energy compaction and denoising performance.
\subsection{Energy compaction}
\begin{figure}[htbp]
    \centering
    \begin{subfigure}{0.48\textwidth}
    \includegraphics[width=\textwidth]{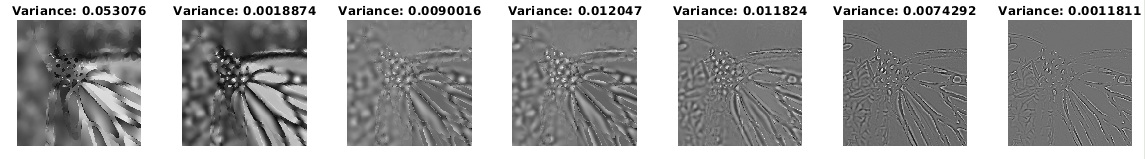}
    \end{subfigure}
    ~~
    \begin{subfigure}{0.48\textwidth}
    \includegraphics[width=\textwidth]{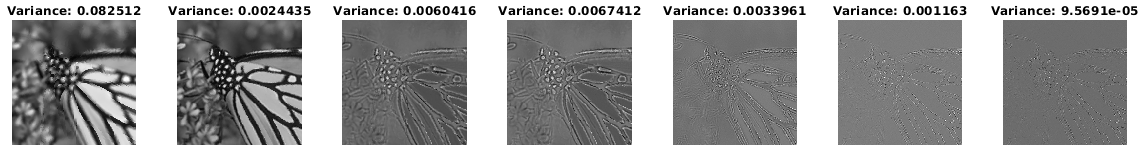}
    \end{subfigure}
    \caption{(Top: $11\times11$ Bilateral Filter Graph vs Bottom: Proposed NNK Graph Construction) Energy compaction using spectral graph wavelets. NNK graphs captures most of image in the lower bands.}
    \label{fig:butterfly_wavelets}
\end{figure}

\begin{figure}[htbp]
    \centering
    \includegraphics[trim={2.5cm 7cm 2cm 6cm},clip,width=0.45\textwidth]{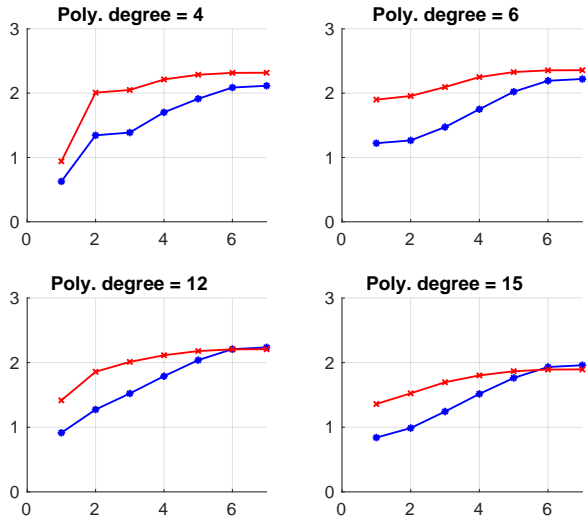}
    \caption{The energy captured by BF Graph (blue) and NNK graph (red) for different polynomial degree approximations of SGW. The wavelets were designed with frame bounds $A=1.71, B=2.35$ as designed in \cite{HammondSGW2011}. NNK consistently captures the image content better than BF graph irrespective of the Chebychev polynomial degree.}
    \label{fig:wavelet_vs_degree}
\end{figure}

In this section, we evaluate our graph construction for image representation. Variance of the wavelet signals is an indicator of information content corresponding to the frequency band of the wavelet. As can be observed in \Figref{fig:butterfly_wavelets}, wavelets corresponding to our method have very less information in the higher frequency bands which is very natural for images as they are inherently smooth. Another feature to identify a good representation of images is the fraction of image energy captured by each wavelet, i.e $(\|\vf_w\|^2/\|\vf\|^2)$. \Figref{fig:wavelet_vs_degree} shows that NNK graphs capture much of the image energy earlier than BF graph which corresponds to compact support in the wavelet domain. 
\subsection{Image Denoising}
\begin{figure}[ht]
    \centering
    \begin{subfigure}{0.4\textwidth}
        \includegraphics[trim={1cm 6cm 2cm 6cm},clip,width=\textwidth]{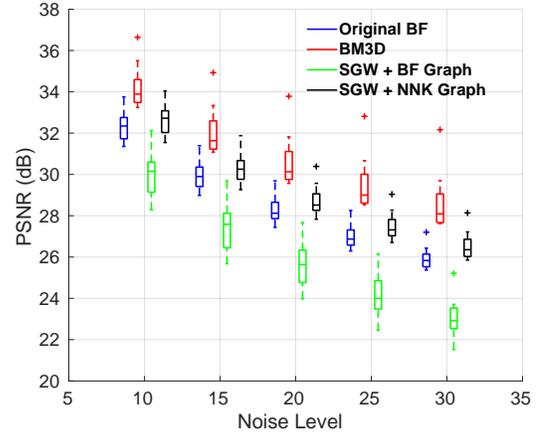}
    \end{subfigure}
    
    \begin{subfigure}{0.4\textwidth}
        \includegraphics[trim={1cm 6cm 2cm 6cm},clip,width=\textwidth]{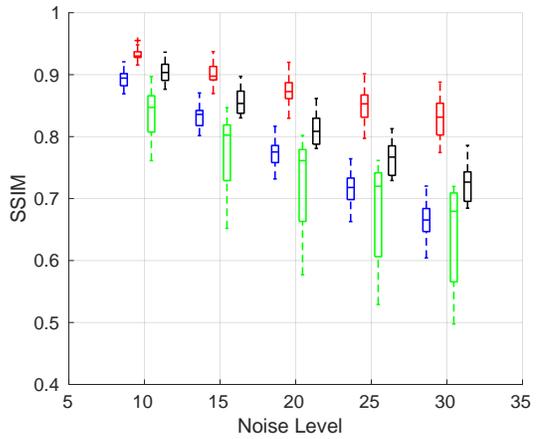}
    \end{subfigure}
    \caption{Denoising performance using SGW on BF graph and proposed method with comparisons to original BF and BM3D algorithms. NNK graphs achieves a significant improvement over the BF Graph version in SSIM and PSNR. Our method improves SSIM of the output while matching PSNR performance with original BF. The BM3D method included for completeness shows that the proposed graph method with BF kernel achieves comparable SSIM measures.}
    \label{fig:denoising_results}
\end{figure}

We consider the problem of image denoising to evaluate filtering performance of our proposed graph. We consider 12 standard images ($256\times 256$) used in image processing 
with Gaussian corruption at 5 different noise variances ($\sigma = 10,15,20,25,30$). We use a $11\times11$ window for constructing the graphs with hyperparameters chosen as in \cite{ZhangMultiBF}. Denoising is done on image signal corresponding to each frequency band separately. The average performance and quantiles for original BF \cite{TomasiBF1998}, BM3D \cite{BM3D_2007} and SGW based on BF graph and proposed method are shown in \Figref{fig:denoising_results}. A key thing to notice is that performance worsens with SGW denoising using BF graph. We attribute this to a shortcoming of the BF graph construction. Since the resulting graph is dense, higher degree polynomials of the BF adjacency used in SGW lead to averaging over larger window and consequently to increased blurring.
\section{CONCLUSION}
We present an attractive framework for image representation using graphs. The proposed graph is sparse and scalable with better energy compaction in its spectral bases than previously used window based graph methods. Our graph construction is robust to a wide range of window sizes and can be run in a parallel for even lower computational complexity.
Further, we explore the use of Spectral Graph Wavelets which operates simultaneously in vertex and spectral domain for image denoising. This approach allows us to leverage ideas from previously studied wavelet methods for images and presents a potential research direction moving forward. In the future, we would like to study the performance of our graph construction with more complex filtering kernels such as those used in non local means, BM3D, kernel regression to mention a few.
\bibliographystyle{IEEEbib}
\bibliography{ICIP_2020}

\end{document}